# Assessing the Performance of a 60-GHz Dense Small-Cell Network Deployment from Ray-Based Simulations

Yoann Corre, Romain Charbonnier, Zahid Aslam
SIRADEL,
Rennes, France

Yves Lostanlen
SIRADEL North America
Toronto, Ontario, Canada
yves.lostanlen@ieee.org

*Abstract*— Future dense small-cell networks are one key 5G candidates to offer outdoor high access data rates, especially in millimeter wave (mmWave) frequency bands. At those frequencies, the free space propagation loss and shadowing (from buildings, vegetation or any kind of obstacles) are far stronger than in the traditional radio cellular spectrum. Therefore, the cell range is expected to be limited to 50 – 100 meters, and directive high gain antennas are required at least for the base stations. This paper investigates the kind of topology that is required to serve a suburban area with a small-cell network operating at 60 GHz and equipped with beam-steering antennas. A real environment is considered to introduce practical deployment and propagation constraints. The analysis relies on Monte-Carlo system simulations with non-full buffer, and ray-based predictions. The ray-tracing techniques are today identified as a relevant solution to capture the main channel properties impacting the beam-steering performance (angular dispersion, inter-link correlation); and the one involved in the present study was specifically enhanced to deal with detailed vegetation modeling. In addition to the user outage, the paper evaluates the evolution of the inter-cell interference along with the user density, and investigates the network behavior in case of local strong obstructions.

## I. INTRODUCTION

Dense small-cell millimeter-wave (mmWave) networks are a promising 5G solution, however many technical challenges like those related but not limited to antennas, signal processing and networking are still to be solved. This paper focuses on the network topology aspect, searching for a small-cell deployment that offers in-street high data rate with almost seamless coverage. Strong propagation attenuation in the mmWave band is obviously a major constraint for the small-cell coverage, which is consequently restricted to few tens of meters. But it also provides high isolation between cells, and thus leads to low interference levels. These are the two sides of a same coin that must be properly taken into account in the network design problem. Determination of the required small-cell density and assessment of the user performance is the objective of the reported work, considering 60-GHz outdoor small-cells. The analysis relies on simulation, where accurate geographical data, propagation ray-based channel modeling, and steerable directive antenna modeling are combined into a DL system-level simulator.

In first part of the paper, the authors propose an extension of the work that was initiated in [1] on mmWave ray-based prediction. The representation of detailed vegetation in the geographical data allows for refined shadowing assessment and leads to more realistic path-loss statistics.

Then in the second part, the propagation model is integrated into a network design exercise. The target is the coverage of main streets and squares into a 300 x 300 m wide surburban area. Small-cells are typically installed on lampposts, with a beamsteering antenna, and EIRP of 40 dBm. Two different antenna beamwidths, 15° and 22°, were tested, expecting that the smaller beamwidth can lead to better user performance (at the price of a more costly antenna alignment process in the operating network). The robustness of the network versus local strong obstruction, i.e. from a bus passing in the street between the small-cell and its users, is also studied. This robustness analysis is only at an early-stage; however it already illustrates one key issue to be managed in mmWave access network performance assessment.

The paper is organized as follows: Section II introduces the ray-based model, which was enhanced to support detailed vegetation representation, then gives some path-loss versus distance statistics in a mixed urban and suburban area. The small-cell deployment scenario under study is described in Section III, as well as the system-level simulation principles. Section IV gives simulation and analysis results, including the proposed network topology and obtained user performance. The network sensitivity to local in-street obstacles is also investigated in this section. Finally, Section V draws some conclusions.

## II. RAY-BASED PREDICTION AT 60 GHZ

The studied ray-based model was introduced in [2] and applied to mmWave case studies in [1]. It predicts the propagation of multi-paths, including the angles of departures (AoD), angles of arrival (AoA) and field strengths, into surburban or urban environments. Specular contributions, i.e.

multi-paths that result from reflections or diffractions, are computed. Diffuse scattering is another allowed contribution, but it was not enabled here. Indeed its characterization at 60 GHz is still an on-going work in the propagation community. As its strength is generally smaller than first-order reflections, we do not expect any major error.

The importance of the environmental details was pointed out in [2] making the use of high-resolution building maps of interest even in theoretical studies (in order to assess realistic shadowing and interference situations) along with the presence of vegetation as a key feature. Propagation through foliage suffers from a loss that basically increases proportionally with the log of frequency [3]. Its impact is severe at mmWave frequency bands with about 11 dB/m at 60 GHz (the value at 60 GHz is obtained from an extrapolation the ITU curve [3]), thus the diffraction above and around the vegetation blocks is the major propagation mechanism compared to the transmission. The woods can be viewed as almost opaque obstacles like buildings, while the isolated trees and hedges need to be carefully considered. The ray-based propagation model predicts two contributions in presence of vegetation: the first one goes through the vegetation; the second one is the diffraction by the vegetation obstacle (using the knife-diffraction method). Only the strongest contribution is preserved. The implemented technique supports the presence of multiple obstacles along each ray path.

Evaluating the impact of vegetation was part of the study in [2], based on a single small-cell simulation, and visual comparison of coverage maps. The study is extended here; a larger number of small-cell locations are considered, and path-loss statistics versus distance are extracted. In addition, an issue in the implementation of the propagation loss through or above multiple vegetation obstacles has been solved.

The study is conducted in a real European environment with a mix of urban and suburban areas. Ten small-cells are created with an antenna height of 6 meters above the ground (typically installed on lampposts). Locations were selected so they are representative of various possible situations: some are placed close to a crossroads; others in the middle of a street; and others in a large square. Predictions run with different vegetation kinds of representation are compared: the first considered geographical map data does not include any information on vegetation; the second one integrates the largest vegetation blocks (surface greater than 300 m²) as sometimes available in high-resolution map data used today for radio-planning; and the last one represents all vegetation details, including the individual trees. The small-cell locations were chosen from the first geographical map data, i.e. without any knowledge on the surrounding vegetation, in order to not bias the evaluation study. The small-cells were moved few meters away from the original position when they appeared to be close to a tree.

The power maps predicted from the small-cell being the most impacted by vegetation are given in Fig. 1. The displayed power is relative to the free-space received power at 1 m; this facilitates the comparison of shadowing effects at different frequencies, i.e. 2.4 GHz and 60 GHz in this particular scenario. The predicted maps clearly illustrate how critical is the presence of vegetation in mmWave band, even isolated trees.

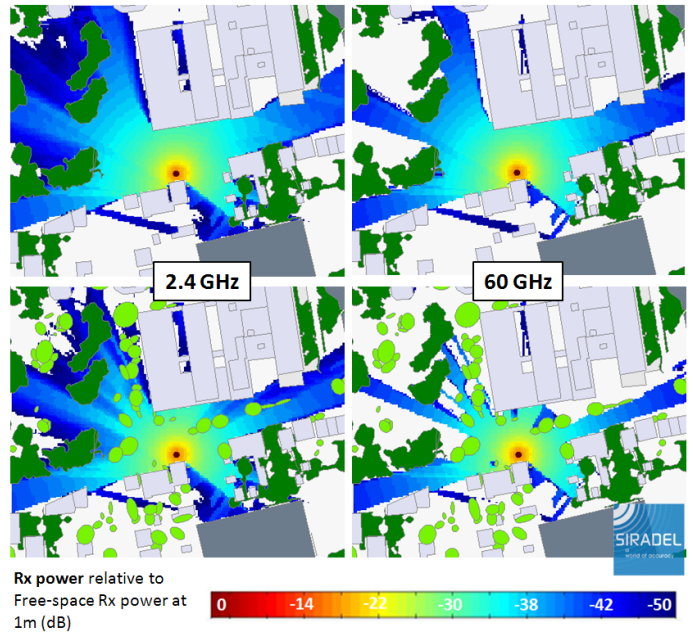

Fig 1. Power map in presence of vegetation at 2.4 and 60 GHz. Top: only main vegetation blocks. Bottom: all detailed vegetation.

The results from all ten small-cells are processed to extract the statistics given in Fig. 2. The median received power and two percentile values (at 10% and 90%) are computed on successive distance intervals of 5 meters (the considered distance is the one measured in the horizontal plane). The graphs show how the prediction of the received power versus distance is changing when introducing vegetation details in the geographical map data. The maximum observed difference on the median received power due to the vegetation modeling is 4 dB at 2.4 GHz (distance = 32.5 m), while it is 23 dB (distance = 37.5 m).

### III. SMALL-CELL NETWORK SIMULATION

The deployment of a dense small-cell network that operates at 60 GHz is evaluated here, aiming to offer outdoor broadband access in a suburban environment. Network performance is assessed from Monte-Carlo simulations, using the detailed geographical map data and propagation ray-based model introduced in the previous section. The results from various user densities and two different antennas are compared in order to assess how the inter-cell interference behaves and affects the link performance. Finally, we simulate a strong obstruction by a bus driving in the street, and observe whether the network can efficiently overcome this trouble or not.

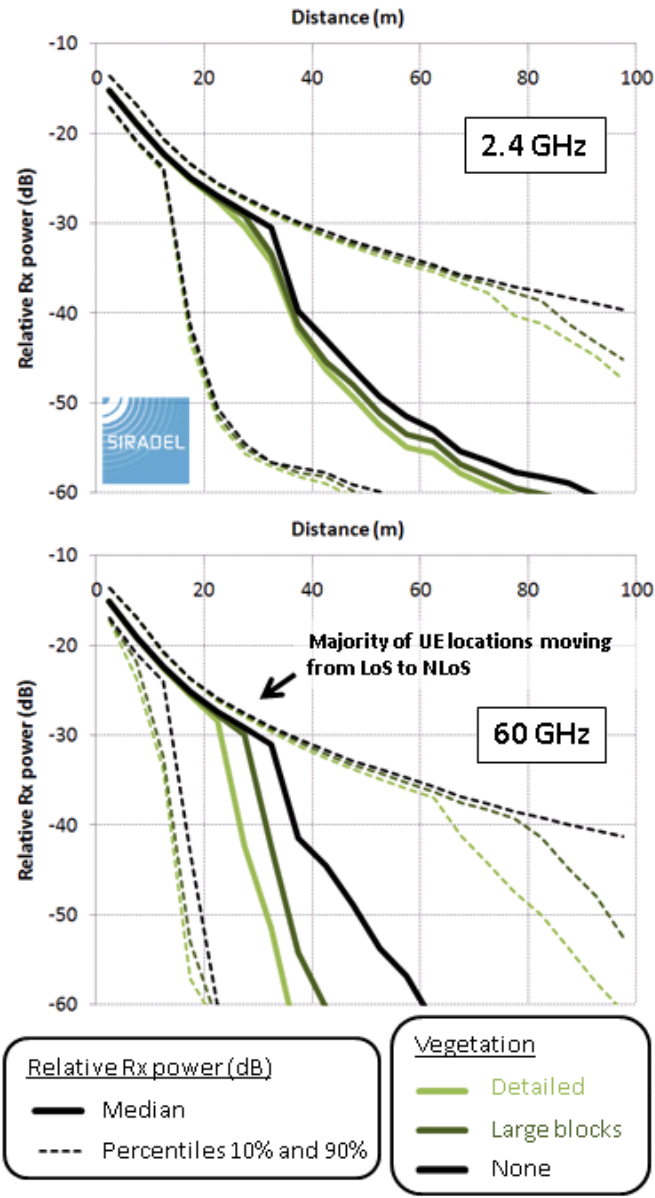

Fig. 2. Received power as a function of distance, considering different vegetation representations and frequencies.

## A. Description of the scenario

The objective of this simulation-based study consists in deploying small-cell antennas so that the main streets and squares in a 300 m × 300 m wide area are covered with broadband mobile access. The small-cells are assumed to be installed on lampposts, i.e. on one side of the street, at height 7 m above the ground. The location of each small-cell must be chosen so it is in line-of-sight from at least one neighbour small-cell, and preferably two. In this way, we make possible the deployment of a wireless mesh network that would likely operate in the mmWave spectrum as well.

Each small-cell has only one sector. It is equipped with an automatically steerable antenna. The half-power beamwidth (HPBW) is 22° with maximum gain 18.5 dBi. The radiation pattern is assumed to be perfectly shaped, i.e. without any significant side-lobe, and following the model given in [4]. The transmit power is adjusted such the EIRP is 40 dBm as allowed in current FCC rules.

Target users are assumed to be all outdoors. It may be objected that some signal penetrates through the windows of buildings in the small-cell close vicinity, however the resulted indoor coverage is very limited and poor; the number of indoor users served in that way must be very small. Relays can obviously be used to combat the strong penetration loss; an antenna installed on the building exterior façade is connected to the outdoor small-cell, while another antenna feeds a local indoor wireless access network. Simulating such system requires the definition a more complex scenario and was out scope of this study. Thus the considered users are all distributed in the streets and open squares. The locations we decided to serve are represented by the blue breaklines in Fig. 3. Remark that courtyards or small streets are not part of the coverage objective, as those areas are difficult to serve and probably with lower traffic demand.

Two active user densities are computed and compared, i.e. 200 users/km² and 1000 users/km², in order to assess the evolution of the inter-cell interference and user performance./ Other scenario parameters are listed in Table I.

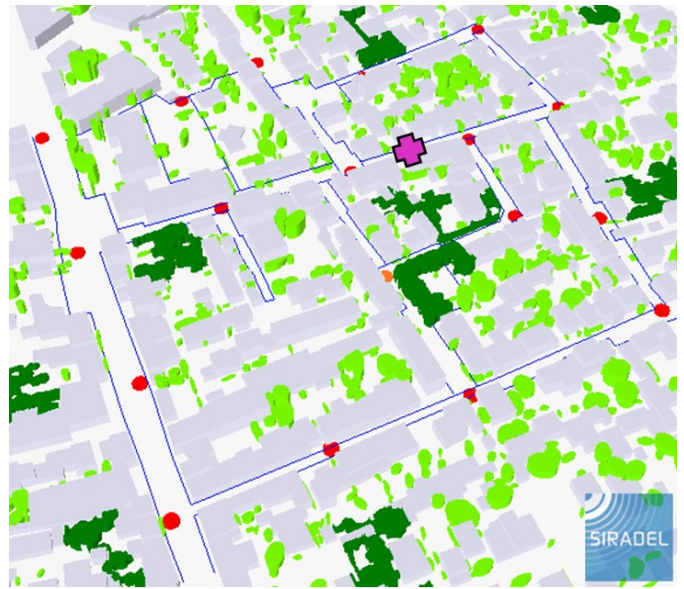

Fig. 3. Target service locations are represented by blue lines; and proposed small-deployment by red dots. The study area is 300 m × 300 m wide.

## B. System-level simulation principles

The simulator is similar to the one run in [6] for analysis of beamforming and beam-switching antennas in a 2-GHz small-cell networks. Users are randomly dropped in the prediction area at successive Monte-Carlo iterations. At each iteration, the simulator determines the attachment and selects the antenna beam orientation that maximizes the downlink (DL) signal-to-noise ratio (SNR) of each user. A procedure similar to the beamforming training specified in IEEE 802.11ad standard [7] permits to find the suitable beam orientation with azimuth resolution 11°, i.e. half the antenna HPBW.

TABLE I. SCENARIO AND SIMULATION PARAMETERS.

| | |
|---|---|
| System | - PHY waveform: Single-Carrier (SC)<br>- Central frequency: 60 GHz<br>- Bandwidth: 200 MHz<br>- Mapping table: from IEEE 802.11ad [5], with 12 adaptative modulation schemes (MCS)<br>- PHY net data rate: 43.8 to 525.0 Mbps |
| Small-cells | - Location: in-street (on lamppost)<br>- Height: 7 meters above ground<br>- Steerable antenna: max gain 18.5 dBi and HPBW 22°<br>- Tx power: 21.5 dBm<br>- Single sector |
| Users | - Location: Outdoor<br>- Height: 2 meters above ground<br>- Noise figure: 7 dB<br>- Antenna: omni-directional, 5 dBi<br>- Demand : 15 Mbps<br>- Active user density: from 200 to 1000 users/km² |
| Power budget | - Impairment loss: 8 dB<br>- DL Rx sensitivity: -79 to -64 dBm |
| Simulation parameters | - Monte-Carlo: 30 iterations<br>- Traffic: Non full-buffer |

The simulator calculates the DL interference levels depending on the antenna beam orientation and bandwidth occupancy by other-cell users. For both the calculation of the useful and interfering powers, the isotropic multi-path field strengths are combined with the antenna gains given in the propagation directions.

Beamforming is realized in the horizontal plane only, meaning that the azimuth of the beam is adjusted for each served user, while the elevation of the maximum gain is always in 0°. A spectrum resource can only used to serve a single user, i.e. multi-user beamforming is not supported. And there is no inter-cell coordination.

Non full-buffer (or finite) traffic is considered. The demand of each active user is limited to 15 Mbps.

IV. SIMULATION RESULTS

The proposed small-cell network is shown in Fig. 3. It has been designed manually and provides close to full coverage on the target service area. It is composed of 18 small-cells over a 0.090 km² wide area. The average inter-site distance (ISD) is 78 m, while the maximum ISD is 110 m. One small-cell, represented by the orange dot in Fig. 3 has been moved up to 12 m above the ground in order to reduce (but not totally remove) the shadowing loss due to surrounding trees. Following subsections analyze in more details the network and user performance depending on the beam-switching antenna, data traffic and local obstruction.

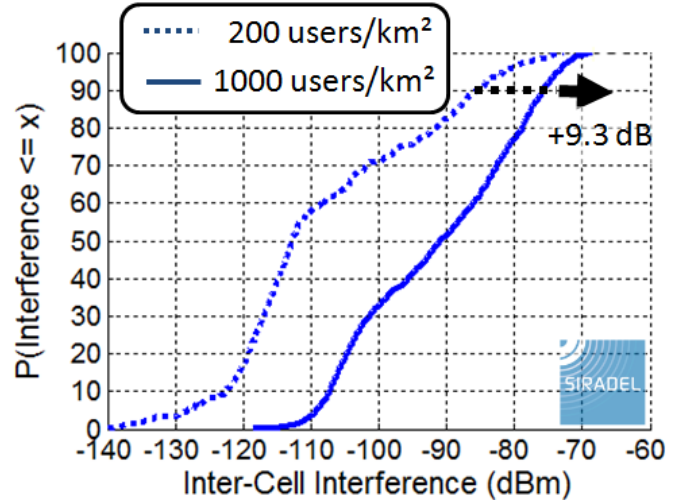

Fig. 4. Evolution of the Inter-cell Interference level vs User density and Antenna.

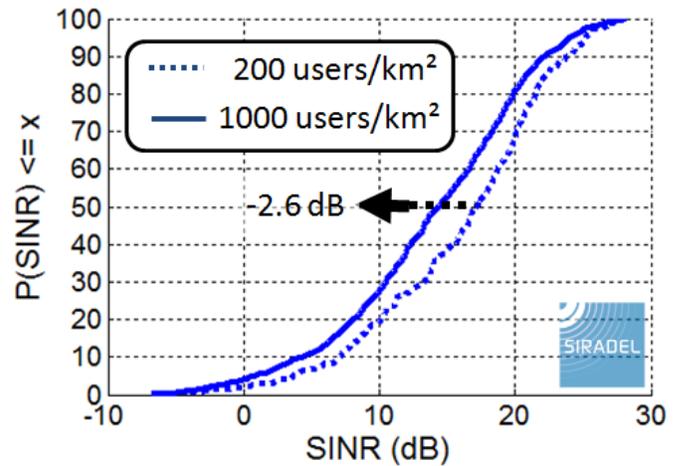

Fig. 5. Evolution of SINR vs User density and Antenna.

A. Inter-cell interference

Fig. 4 and Fig. 5 show respectively the cumulative distribution function (CDF) of the inter-cell interference levels and signal-to-interference-plus-noise-ratio (SINR) for two different user densities. Except for a very small occurrence percentage, the interference level with 200 users/km² is below and even far below the receiver sensitivity (-79 dBm), meaning that the network performance is almost not affected by the interference. When going up to 1000 users/km², the percentage of allocated resources by each small-cell strongly increase (the average cell load goes from 4.1% to 22.6%, and the load in the most crowded cell goes from 7.6% to 46.4%). Thus the interference levels grow significantly; the percentage of them above the receiver sensitivity goes up to 20%. The peak interference value, measured at the 90-percentile, increases from -85 dBm to -76 dBm. Consequently, the SINR perceived by the users is degraded. Only in a quite limited way however: the mean SINR decreases by 2.6 dB, which is reasonable knowing that the amount of allocated resources in the network has been multiplied by a factor of 5.5. Actually the strong propagation loss and shadowing, as well as the beamsteering-based spatial filtering, permit to efficiently isolate users located in two different cells.

The same analysis was conducted with another antenna, having a smaller beamwidth (15°), a steering azimuth resolution of 6°, and a maximum gain of 21.9 dBi. The 40 dBm EIRP is maintained (maximum radiated power currently allowed by FCC for 60-GHz for point-to-point links), meaning that the transmit power has been reduced from 21.5 dBm to 18.1 dBm. As EIRP is kept constant, a few dB decrease of both the received and interfering signal strengths is observed, which comes from a reduction in the multi-path combination gain; the decrease on interfering levels is significantly greater than the one on the useful signal in average. All this was expected. An improvement in SINR levels was expected as well; the inverse trend is observed. The reason is that most users suffer from the signal strength degradation, while only a small percentage of links were interference-limited, thus benefit from the interference decrease. The conclusion is that: if the antenna gain difference is compensated by a transmit power adjustment, and the global interference strength suffered by the network is low, then a smaller beamwidth does not necessarily lead to better performance.

*B. User performance*

The user quality of service has been measured from three metrics: user outage, mean DL spectral efficiency (SE), and the mean DL spectral efficiency at cell-edge, which is the average spectral efficiency computed from the 10% worst-served user. Fig. 6 gives the evolution of those metrics depending on the user density. Outage is twice when increasing the user traffic. The mean SE and mean cell-edge SE are respectively degraded by 3% and 13%.

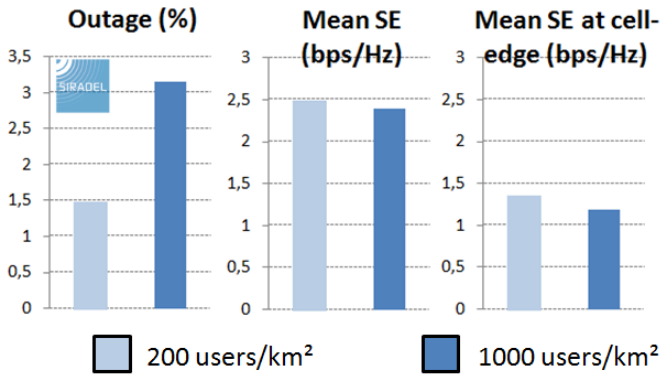

Fig. 6. User performance vs User density and Antenna.

*C. Strong local obstruction*

Because of the small-cell height and transmission frequency, the coverage may be strongly affected by in-street obstructions, like static objects (e.g. urban furniture), human bodies or high vehicles. Actually the obstruction of a dominant path can be partly mitigated by the system by pointing the antenna towards another path. This mechanism can be properly predicted if multi-path channel data is available. As an example, in Fig. 7, the bus is obstructing the dominant direct-path, but significant signal strength still reach the user location from a reflection on the building façade.

In the considered obstruction scenario, a bus is located at the pink cross shown in Fig. 3. The network performance is computed for the users located in the same street, between both small-cells surrounding the bus. Inter-cell interference is neglected. Finally, the simulation leads to a full user coverage even in close vicinity of the bus (no user outage) and only a 2.4 dB degradation on the cell-edge DL SINR. *The. analysis will be enhanced in the final paper version with more complete and representative statistics.*

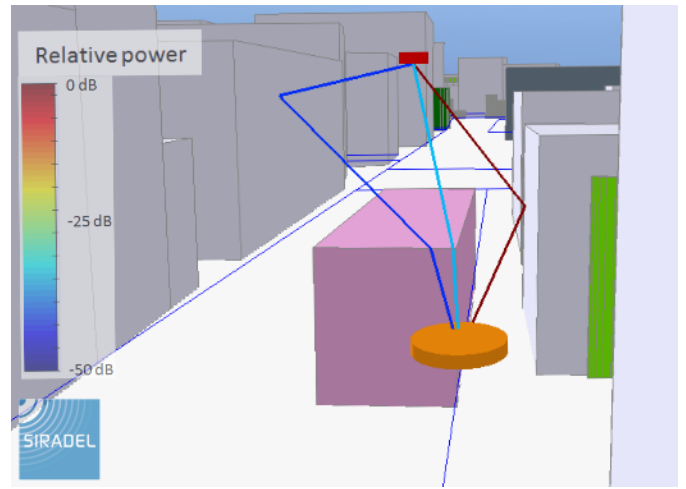

Fig. 7. Impact of a bus obstruction on the propagation paths.

V. CONCLUSION

The topology of future outdoor mmWave small-cell networks must adapt to various constraints like the propagation loss, presence of opaque or quasi-opaque static obstacles in the streets (vegetation, urban furniture), and high sensibility to in-street moving obstacles (vehicles, bodies). Accurate geographical map data and an enhanced ray-based model are coupled here with a Monte-Carlo system-level simulator to determine what small-cell density is required to reach an almost full coverage in main streets of a suburban environment (18 small-cells over 0.09 km² wide area). The impact of the steerable antenna beamwidth and user density on the inter-cell interference and user performance are assessed as well as the degradation suffered by a bus obstruction. Good robustness of the network versus interference and in-street obstruction was observed.

## REFERENCES

[1] Y Corre, T Tenoux, J. Stéphan, F. Letourneux and Y Lostanlen, "Analysis of Outdoor Propagation and Multi-Cell Coverage from Ray-Based Simulations in sub-6GHz and mmWave Bands," in 10th European Conference on Antennas and Propagation (EuCAP), 2016.

[2] Y. Corre and Y. Lostanlen, "Three-dimensional urban EM wave propagation model for radio network planning and optimization over large areas," *IEEE Transactions on Vehicular Technology*, vol. 58, no. 7, pp. 3112-3123, 2009.

[3] ITU-R Rec. P.833-7, *Attenuation in vegetation*, 2012.

[4] A. Maltsev, et al., *Channel Models for 60 GHz WLAN Systems*, IEEE document 802.11-09/0334r8, May 2010.

[5] IEEE Std 802.11ad, *Wireless LAN Medium Access Control (MAC) and Physical Layer (PHY) Specifications, Amendment 3: Enhancements for Very High Throughput in the 60 GHz Band*, 2012.

[6] G. Gougeon, Y. Corre, A. De Domenico, A. Clemente, A. Sattar Kaddour, S. Bories and Y. Lostanlen, "LTE System-Level Evaluation of Directive Compact Antennas for Small-Cell Networks," in 10th European Conference on Antennas and Propagation (EuCAP), 2016.

[7] T. Nitsche, C. Cordeiro, A. B. Flores, E. W. Knightly, E. Perahia, J. C. Widmer, "IEEE 802.11ad: directional 60 GHz communication for multi-Giga-bit-per-second Wi-Fi", *IEEE Communications Magazine*, Vol. 52, Issue 12, Dec. 2014.